\documentclass[prd,aps,twocolumn,nofootinbib]{revtex4}
\usepackage[utf8]{inputenc}
\usepackage{graphicx}
\usepackage{xcolor}
\def\gr{$\gamma$-ray}
\def\tibet{Tibet-AS$\gamma$}

\begin{document}

\title{Pion decay model of TIBET-AS$\gamma$ PeV  gamma-ray signal}
\author {Sergey Koldobskiy$^{1,2}$, Andrii Neronov$^{3,4}$,  Dmitri Semikoz$^{3,5,1}$, }
\affiliation{$^1$National Research Nuclear University MEPhI, 115409 Moscow, Russia\\
$^2$University of Oulu, 90570 Oulu, Finland\\
$^3$Université de Paris, CNRS, Astroparticule et Cosmologie,  F-75006 Paris, France\\
$^4$Astronomy Department, University of Geneva, Ch. d'Ecogia 16, 1290, Versoix, Switzerland\\
$^5$Institute for Nuclear Research of the Russian Academy of Sciences, Moscow, Russia \\
}

\begin{abstract}
\tibet\ collaboration has recently reported a measurement of diffuse \gr\ flux from the outer Galactic disk in the energy range reaching PeV.  We complement this measurement with the Fermi/LAT measurement of the diffuse flux from the same sky region and study the pion decay model of the combined Fermi/LAT$+$\tibet\ spectrum. We find that within such a model the average cosmic ray spectrum in the outer Galactic disk has the same characteristic features as the local cosmic ray spectrum. In particular, it experiences a hardening at several hundred GV rigidity and a knee feature in the PV rigidity range. The slope of the average cosmic ray spectrum above the break is close to the locally observed slope of the helium spectrum $\gamma\simeq 2.5$, but is harder than the slope of the local proton spectrum in the same rigidity range. Although the combination of Fermi/LAT and \tibet\ data points to the presence of the knee in the average cosmic ray spectrum, the quality of the data is not yet sufficient for the study of knee shape and cosmic ray composition.  
\end{abstract}

\maketitle

\section{Introduction}

\tibet\ collaboration has recently reported a measurement of diffuse \gr\ flux from the outer Galactic Plane up to the PeV energy range \cite{Amenomori2021}. This measurement possibly provides information about cosmic rays with energies above PeV residing in the interstellar medium across the Milky Way disk. In this sense it potentially carries information complementary to that obtained using the local measurements of the cosmic ray flux at the location of the Earth. 

\begin{figure}
\includegraphics[width=\columnwidth]{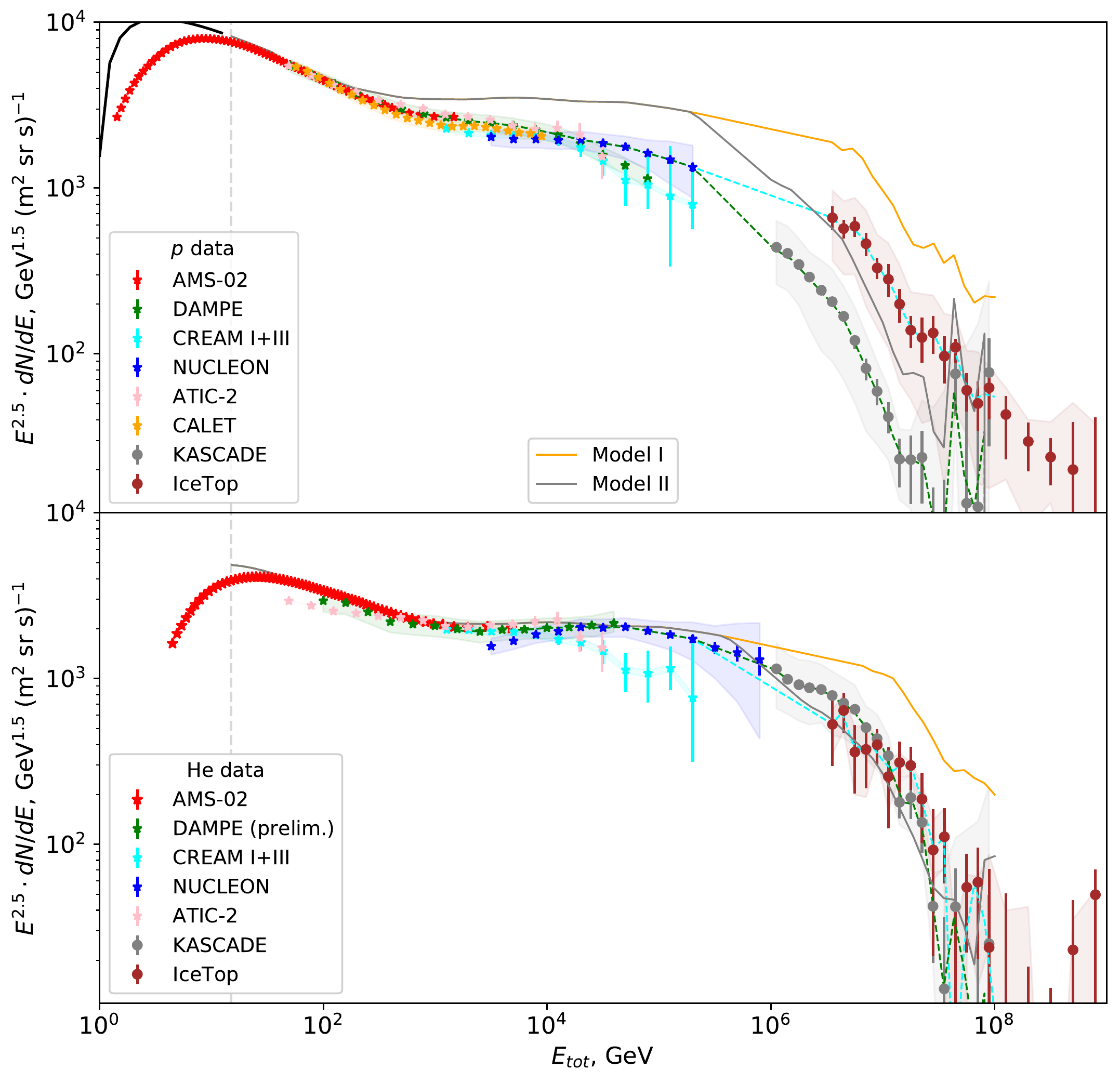}
\caption{Summary of local measurements of the cosmic ray proton (top) and Helium (bottom) fluxes by AMS-02 \cite{Aguilar2020a}, CREAM \cite{Yoon2017}, ATIC-2 \cite{Panov2009}, DAMPE \cite{An2019, Alemanno2021}, CALET \cite{Adriani:2019aft}, NUCLEON \cite{Grebenyuk2019}, IceTop \cite{Aartsen2019}, and KASCADE \cite{Apel2013} experiments. Cyan and green dashed curves show interpolation of the proton and helium data points passing through, respectively, IceTop and KASCADE data. Orange and grey curves in the top panel show the modified model proton spectra discussed in the text. For comparisons these modified proton spectra are shown in the bottom panel as well.  {Spectra are given in the total energy units.}}
\label{fig:CR_spectrum}
\end{figure}

The local cosmic ray spectrum measurements from GeV to multi-PeV energy range obtained by AMS-02 \cite{Aguilar2020a}, CREAM \cite{Yoon2017}, ATIC-2 \cite{Panov2009}, DAMPE \cite{An2019, Alemanno2021}, CALET \cite{Adriani:2019aft}, NUCLEON \cite{Grebenyuk2019}, IceTop \cite{Aartsen2019}, KASCADE \cite{Apel2013} reveal a range of puzzling properties of the spectrum. Overall, the spectra of different components of cosmic ray flux are broken powerlaws  $dN/dE\propto E^{-\gamma}$  {(here we use conventional symbol $E$ for the total energy of the particle)} in the energy range much below the knee at $E_\mathrm{knee}\simeq 4$~PeV \cite{Antoni:2005wq,Amenomori:2008aa,Abbasi:2018xsn}.  The spectra of different nuclei all exhibit a hardening break at several hundred GV rigidity \cite{2011Sci...332...69A}. The slope of the proton spectrum is changing from $\gamma_{p1}\simeq 2.87$ down to $\gamma_{p2}\simeq 2.56$ \cite{2011Sci...332...69A,Adriani:2019aft} while the slope of the helium spectrum is hardening to $\gamma_{\mathrm{He}2}\simeq 2.5$ \cite{Grebenyuk2019}, see Fig.~\ref{fig:CR_spectrum}. 

The difference in the slopes of different flux components is puzzling.  Cosmic ray acceleration and propagation models typically evoke physical processes that universally scale with particle rigidity. Changes in the spectra of different nuclei are expected to be the same if the spectra are expressed as functions of rigidity. 
It is not clear if the hardening in cosmic ray spectrum and difference of slopes of proton and helium (and other primary cosmic ray nuclei) spectra are a local cosmic ray feature or they are a generic properties of the Galactic cosmic ray spectrum.

Equally uncertain is the origin  of the knee of the cosmic ray spectrum (see \cite{Kachelriess:2019oqu} for a recent review). It can correspond to the highest energy of cosmic rays produced by Galactic sources \cite{end_of_galactic,Stanev:1993tx} or to the break in the cosmic ray spectrum of those sources
\cite{Drury:2003fd,Cardillo:2015zda} or it  can occur at the energy at which particle Larmor radius is comparable to the correlation length of Galactic magnetic field in "escape" model \cite{escape,Giacinti:2014xya,Giacinti:2015hva}. 
The energy of the knee and the details of the spectral changes at the knee might also be different across the Galaxy  if  the knee is locally dominated by one or few sources \cite{erlykin,Bouyahiaoui:2018lew}. This possibility is natural in the anisotropic diffusion model which brings phenomenological diffusion models in agreement with the realistic $\mu G$ scale magnetic field of Galaxy \cite{Giacinti:2017dgt}. 

Measurements of the energies of the knees of different flux components disagree with each other. KASCADE experiment using QGSJET-II-02 model has found the proton knee at the energy about $E_p\simeq 2$~PeV, while the knee in the helium spectrum is at $E_\mathrm{He}\simeq 4$~PeV, so that the two knees occur at the same rigidity \cite{Apel2013}. ARGO-YBJ  measurements suggest that the proton knee is at  lower energy $E_p<1$~PeV \cite{argo}. To the contrary, IceTop experiment using Sibyll 2.1 model finds the knees of the proton and helium spectra at approximately the same energy $E_{p-\mathrm{He}}\simeq 4$~PeV \cite{Aartsen2019}. 
Since all-particle spectrum is the same in all experiments, the origin of these discrepancies is in large systematic uncertainty of composition reconstruction in data analysis of different experiments, both in extraction of many nuclei groups from the data and in dependence of results on hadronic interaction models (see \cite{lipari_knee} for an overview). 

Diffuse \gr\ flux measurements from the Galactic disk region provide a possibility to constrain the properties of the average cosmic ray spectrum in the disk \cite{Neronov:2011wi,neronov15,2016PhRvD..93l3007Y,2017A&A...606A..22N,Neronov:2020wir}. In general, the diffuse \gr\ flux is composed of several components, including Bremsstrahlung and inverse Compton emission from cosmic ray electrons and neutral pion decay emission from interactions of cosmic ray protons and nuclei with interstellar medium \cite{fermi_diffuse12,fermi_models,Lipari2018}. However, the pion decay component is strongly boosted, compared to the electron component of the \gr\ flux, in the densest part of the Galactic disk (a region of about $\pm 150$~pc around the Galactic Plane) \cite{Lipari2018}. This suggests that the spectral template of the pion decay emission component can be isolated by subtracting the diffuse flux at Galactic latitude above certain Galactic latitude cut (e.g. $5^\circ$) from the flux measurements below this cut. In this way, the \gr\ flux components sensitive to the density of the interstellar medium (Bremsstrahlung and pion decay) would be boosted compared to the inverse Compton component \cite{neronov15}. 

Such an approach has been chosen in \tibet\ data analysis \cite{Amenomori2021}, in which the sky region $|b|>20^\circ$ has been chosen as "background estimate" region and the region at $|b|<5^\circ$ has been considered as the "signal" region.

In what follows we implement the approach of \tibet\ data analysis for the analysis of Fermi/LAT data, to complement \tibet\ measurements with lower energy data points. In this way we obtain the spectral template for the pion decay $+$ Bremsstrahlung emission from the outer Galactic Plane (Galactic longitude range $50^\circ<l<200^\circ$). We model the resulting diffuse \gr\ spectrum in a broad GeV-PeV energy range using a "minimal" pion-decay-only model, to get a first idea on possible range of properties of the average cosmic ray spectrum in the outer Galactic disk. 


\section{Fermi/LAT data analysis}

Our analysis of Fermi/LAT data adopts the same approach as \tibet\ data analysis \cite{Amenomori2021}. We consider diffuse flux from the sky region $50^\circ<l<200^\circ$, $|b|<5^\circ$. To get rid of the isolated source flux, we remove photons from within circles of the radius $0.5^\circ$ around sources from the 4th Fermi/LAT catalog (option 1) or around the sources from TeVCat online catalog of TeV \gr\ sources (option 2). We use Pass 8 Fermi/LAT dataset spanning 12 years (2008--2020), the {\tt SOURCEVETO} event selection. To calculate the diffuse source flux we collect events in square boxes of $5^\circ$ filling the region of interest. In each box and each energy bin, we calculate the exposure using the {\it gtmktime-gtexposure} Fermi Science Tools routine combination\footnote{\url{https://fermi.gsfc.nasa.gov/ssc/data/analysis/software/}}. 

Following the approach of \tibet \cite{Amenomori2021}, we use photon counts in the sky region $|b|>20^\circ$ to estimate the "background" counts in each energy bin and each $5^\circ$ box. As discussed in the Introduction, this "background" flux in fact contains the diffuse emission flux at higher Galactic latitude, which has higher inverse Compton flux component. In this way, the analysis "boosts" the pion decay and Bremsstrahlung components and suppresses the inverse Compton component. 

\begin{figure}
\includegraphics[width=\columnwidth]{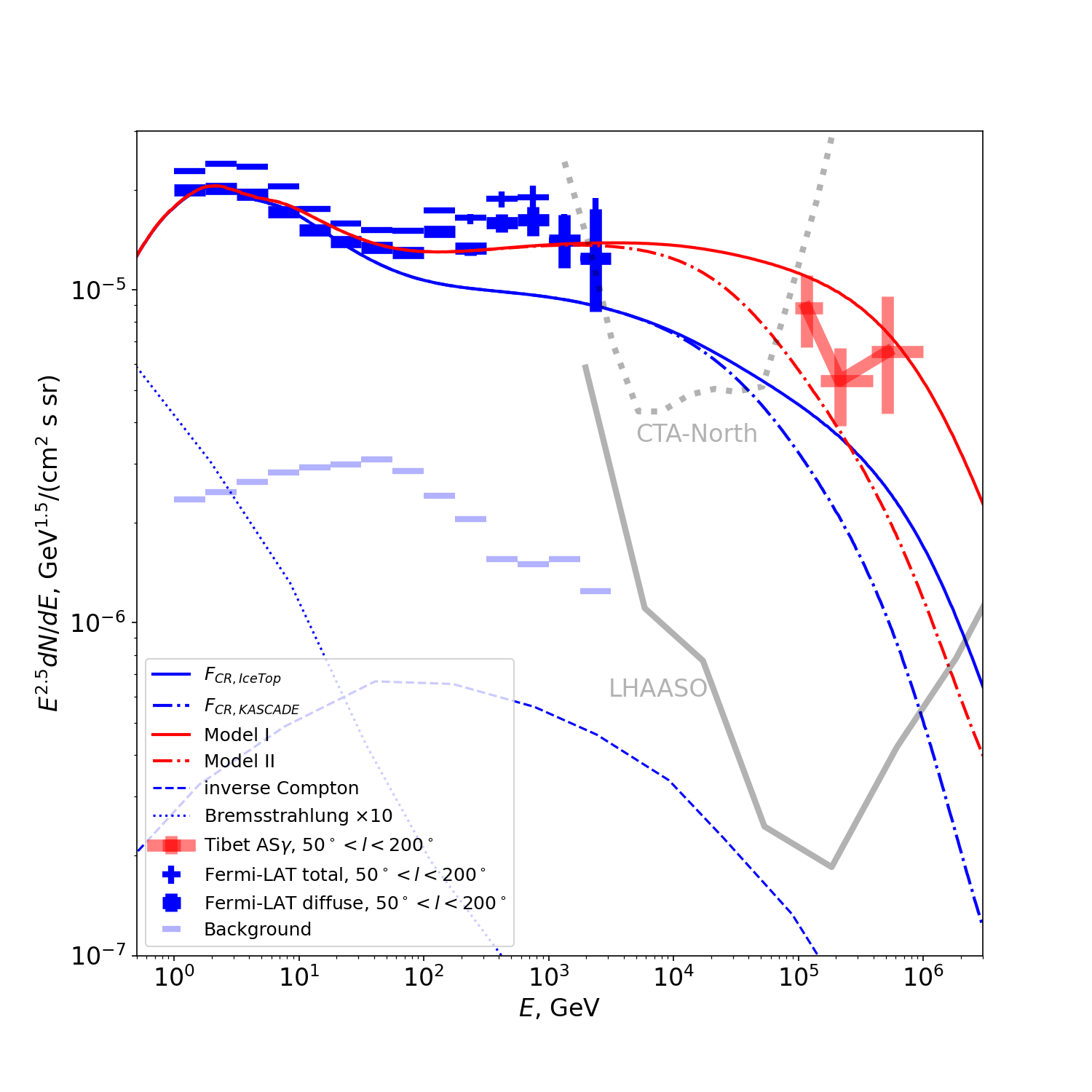}
\caption{ Gamma-ray flux from the $50^\circ<l<200^\circ, |b|<5^\circ$ region of the outer Galaxy.   
Total Fermi/LAT  gamma-ray flux measurements are  shown with thin blue data points, the spectrum of diffuse emission component of the flux  shown with thick blue data points.  {Background level is shown by light-blue color.}  \tibet \, measurements \cite{Amenomori2021} are shown with red data points.   {Dash-dotted and solid blue lines} show model pion decay spectra calculated for the locally measured cosmic ray spectra with composition corresponding to KASCADE and IceTop measurements.  {Dash-dotted and solid red} lines show the \gr\ flux for the Models I and II of the proton spectrum consistent with local measurements of He spectrum, as described in the text.  {For comparison, inverse Compton and Bremsstrahlung flux levels corresponding to the local interstellar medium emissivity \cite{Lipari2018} are shown by blue dashed and dotted lines}. Thick solid and dashed grey lines show sensitivities of LHAASO \cite{Neronov:2020wir} and CTA-North \cite{Neronov:2020zhd} for diffuse \gr\ flux.  }
\label{fig:gamma_spectrum}
\end{figure}

The resulting \gr\ spectrum is shown in Fig.~\ref{fig:gamma_spectrum}. Pale blue data points show the total flux measurement (or, more precisely, a slight under-estimate of the total flux, related to the specific background estimation procedure \cite{neronov_diffuse}).  Full blue data points show the diffuse emission spectrum after subtraction of the isolated source contribution using Option 1. Subtraction using Option 2 gives a similar result. The spectrum exhibits a noticeable hardening above approximately 30~GeV energy. In this energy range the spectrum is consistent with a power-law with the slope $\simeq 2.5$, as clear from the $E^{2.5}dN/dE$ plot representation in Fig.~\ref{fig:gamma_spectrum} in which the $E^{-2.5}$ spectrum appears as a horizontal line. One can also notice that the \tibet\ measurements in the same representation fall below the extrapolation of the $E^{-2.5}$ power-law to the PeV band. 

\section{Modelling}

The simplest model of the combined Fermi/LAT and \tibet\ diffuse \gr\ flux spectrum shown in Fig.~\ref{fig:gamma_spectrum} is that of \gr\ emission from neutral pion decays. As it is discussed above, the spectral extraction method used by \tibet\ and implemented in our analysis for Fermi/LAT data boosts the pion decay component with respect to the inverse Compton flux component that might also contribute to the flux. However, already the non-boosted pion decay component is expected to largely dominate the inverse Compton component for the flux from the direction of the outer Galactic Plane  \cite{fermi_diffuse12,Lipari2018}.  The Bremsstrahlung component gives only a minor contribution to the flux and only in the GeV energy range and can also be neglected in the first approximation \cite{fermi_diffuse12}.  {This is illustrated by the blue lines in Fig. \ref{fig:gamma_spectrum}, where we plot the pion decay, inverse Compton and Bremsstrahlung spectra corresponding to the local interstellar medium emissivity from Ref. \cite{Lipari2018}. The Solar system is located in the outer Galactic disk and the local emissivity can be considered representative for the emissivity across the outer Galactic disk.  If the pion decay spectrum normalisation is fixed to match the Fermi/LAT data, the inverse Compton contribution to the flux is expected to be at a several percent level in the energy range of interest.}

To model the pion decay flux from different cosmic ray nuclei we use AAfrag package \cite{Kachelriess2019a} based on the QGSJET-II-04m model. We combine AAfrag calculations in the energy range  {above 15 GeV with low energy  parameterisations of Kamae et al.~2006~\cite{Kamae2006}} to obtain the \gr\ spectrum for the full range of cosmic ray energies (the boundary between two models is shown with dashed gray line in Fig.~\ref{fig:CR_spectrum}). 
 {In our calculations we propose that the interstellar medium consists of 91\% of hydrogen and 9\% of helium.}
AAFrag package allows to calculate differential cross-section of \gr\  production separately for different cosmic-ray species while in the model of  {Kamae et al.~2006} only $pp$ interactions are considered. To account for  {\textit{p}He, He\textit{p} and HeHe interactions at low energies, we
calculated and applied an additional nuclear factor of 1.4 to the results of \gr\ production using Kamae et al. code for primary $p$ energy from the threshold energy to 15 GeV}. 

For our reference model we take local measurements of cosmic ray flux  (proton and helium component measurements and our interpolation are summarized in Fig.~\ref{fig:CR_spectrum}). We complement the local flux measurements with a model of Local Interstellar spectrum (LIS) below 40 GeV to take into account the solar modulation effect \cite{Vos2015}. This model explains also the Voyager experimental data  \cite{Stone2013}.

It is important to notice that the local cosmic ray flux component measurements suffer from systematic uncertainties both in the energy range between the sub-TeV break and the knee and in the knee energy range (see Fig.~\ref{fig:CR_spectrum}). The discrepancy in the measurements of different experiments reaches a factor of 2 in the 100 TeV range (it is accounted within either systematic or statistical uncertainty, depending on experiment). In the knee energy range, even though the total flux measurements of different experiments agree with each other, measurements of the spectra of different nuclei are inconsistent. This is illustrated in the top panel of Fig.~\ref{fig:CR_spectrum} in which one can see that KASCADE and IceTop measurements of the proton knee are inconsistent with each other. 

Assuming that measurements of all experiments agree at approximately TeV energy (see Fig.~\ref{fig:CR_spectrum}), a factor-of-two difference in measurements of different experiments at 100 TeV corresponds to a $\delta\gamma\simeq 0.15$ uncertainty of the slope of the spectra of cosmic ray protons and nuclei in the 1--100 TeV energy range. 

These uncertainties of the local cosmic ray spectrum measurements propagate into uncertainties of the pion decay \gr\ flux model. One of these uncertainties, of the energy of the proton knee, is illustrated in the model curves in Fig.~\ref{fig:gamma_spectrum}. In this figure, the two dashed lines show models based on the local proton knee spectra from KASCADE and IceTop. One can clearly see that the two \gr\ flux models strongly differ exactly in the energy range of \tibet. This indicates that the energy range covered by \tibet\ (and LHAASO) is crucially important for exploration of the nature of the knee with \gr\ data. 

Both models, based on KASCADE and IceTop measurements, under-predict the \gr\ flux, compared to the \tibet\ measurement. However, as it is discussed above, the measurements of the local cosmic ray flux in the TeV-PeV range suffer from systematic uncertainties. The estimate of  the pion decay flux in the PeV range vary depending on assumptions about  the energy of the proton (and other nuclei) knee and on the slope of the spectra of cosmic ray flux components in 1--100 TeV range.  It is thus worth exploring what kind of  modifications of the cosmic ray spectral components can make the pion decay model consistent with the data.



All the cosmic ray nuclei spectra have a break at several hundred GV rigidity. The simplest possibility (theoretically motivated) is that the slopes of the average cosmic ray nuclei spectra above the break are the same. We consider this possibility by modifying the slope of the proton spectrum by $\delta\gamma=0.1$ above a break at 100 GeV, by multiplying the proton flux  $F_p$ by a broken powerlaw
\begin{equation}
    F_{p,\mathrm{mod}}(E)=F_p\left(1+\left(\frac{E}{100\mbox{ GeV}}\right)^2\right)^{0.1/2}
\end{equation}
After such a modification, the proton spectral models in the 1--100 TeV range match the shape of the helium spectrum, while the KASCADE and IceTop proton knees bracket the errorbars of the helium knee measurements with both KASCADE and IceTop, see Fig. \ref{fig:CR_spectrum}. We call Model I the "hardened" proton spectrum with the knee shape based on KASCADE data and Model II the model in which the knee shape is based on the IceTop data.

This modification of the proton spectrum increases the strength of the break in the \gr\ spectrum at approximately 1--30 GeV energy and makes the pion decay spectrum in the energy range above 30 GeV consistent with Fermi/LAT measurements (see Fig.~\ref{fig:gamma_spectrum}). The model solid lines in Fig.~\ref{fig:gamma_spectrum} show the pion decay flux in the modified proton spectrum model for the two knee models based on KASCADE and IceTop data.  

One can see that after such minor modification, the model \gr\ spectra for both KASCADE-based and IceTop-based composition models provide a satisfactory description of the joint Fermi/LAT $+$ \tibet\  \gr\ spectrum. 

\section{Discussion}

The analysis presented above allows us to formulate a hypothesis that the average cosmic ray proton and helium spectra in the outer Galactic disk have identical shape in the TV-PV rigidity range, with the average proton spectrum being somewhat harder than the locally measured proton spectrum, but consistent with local helium spectrum. Within such a hypothesis, the spectrum of diffuse \gr\ emission from the dense part of the Galactic disk in the longitude range $50^\circ<l<200^\circ$ is well described by the pion decay model over the entire GeV-PeV range covered by Fermi/LAT and \tibet. 

An attractive property of this hypothesis is that the identical shapes of the proton and helium spectra are expected on theoretical grounds. Both the acceleration and propagation physical mechanisms are typically sensitive for particle rigidity and hence the spectra of different cosmic ray nuclei coming from certain source type are expected to be identical as a function of rigidity. 

Indication of harder average slope of the cosmic ray spectrum with the slope $2.5$ in the Galactic disk have been previously found in Fermi/LAT data \cite{neronov15,2016PhRvD..93l3007Y,fermi_models,kra-gamma}. The results presented here reveal additional evidence for this possibility and add new details. Fits to the spectra of diffuse \gr\ emission from the inner Galactic disk indicate that in this part of the Galaxy the data are consistent with the cosmic ray spectrum slope $2.5$ all over the GV-PV energy range. This is not the case locally and it seems to be not the case for the cosmic ray spectrum in the outer Galactic disk. Instead, in the outer Galactic disk the cosmic ray spectrum hardens to the 2.5 slope only above several hundred GV rigidity. 

The slope of the TV-PV cosmic ray spectrum is consistent with the model in which cosmic rays are accelerated by their sources with e.g. $dN/dE\propto 1/E^{2.2}$ spectrum produced by shock acceleration and propagate in Galactic magnetic field with Kolmogorov turbulence, which will soften spectrum to $dN/dE\propto 1/E^{2.5}$. The spectral break at several hundred GV rigidity can be explained by several mechanisms: cosmic-ray induced turbulence  \cite{Blasi:2012yr,Aloisio:2013tda}, two-component halo model \cite{Tomassetti:2015mha} or by two or more source populations model 
\cite{Zatsepin:2006ci,Thoudam_2012}. The discrepancy between locally measured and the average cosmic ray proton spectrum with the $\sim 2.5$ slope at $E> 1$ TeV can be a local feature. This feature arises in anisotropic diffusion scenario \cite{Giacinti:2017dgt} in which small number of sources   dominate the local cosmic ray flux in multi-TV range \cite{Kachelriess:2017yzq,Bouyahiaoui:2018lew,Bouyahiaoui:2020rkf}. 

Better information on variations of the spectrum of cosmic rays across the Galactic disk can be obtained with higher statistics measurements of the variations of the spectrum of diffuse emission from the Galactic disk as a function of Galactic longitude. Closing the gap in the measurements in 3--100 TeV energy range would also provide better quality measurement of the slope of the average cosmic ray spectrum in TV-PV rigidity range.
This will be possible with the next generation \gr\ telescopes CTA and LHAASO. Fig.~\ref{fig:gamma_spectrum} shows the sensitivities of CTA-North \cite{Neronov:2020zhd} and LHAASO \cite{Neronov:2020wir} to the diffuse \gr\ flux from the Galactic Plane. CTA-North will be able to close the 3--100 TeV gap and provide better separation between the truly diffuse and isolated source contributions to the \gr\ flux in the  part of the Galactic Plane covered by \tibet\ measurement. LHAASO will provide high-statistics measurements all the way up to the PeV range. This will allow to constrain the shape of the cosmic ray spectrum at the knee and to distinguish between different theoretical models of the knee: maximum energy of Galactic sources,  influence of a single source, change of propagation regime.

If the diffuse \gr\ emission is produced by the pion decays, it is accompanied by the neutrino flux. The \tibet\ measurement of the diffuse flux from the outer Galactic Plane is comparable to the IceCube measurement of the sky-average astrophysical neutrino flux \cite{icecube}. This suggests that the neutrino counterpart of the \tibet\ \gr\ flux measurement should be detectable with deeper IceCube exposure (currently a p-value $\simeq 0.02$ is found for a specific template of the all-Galactic emission that includes the Galactic Plane \cite{icecube_kragamma}) and with Baikal-GVD \cite{baikal-gvd} and KM3Net \cite{km3net} neutrino telescopes. This makes the pion decay model of the diffuse \gr\ flux from the sky region $50^\circ <l<200^\circ$, $|b|<5^\circ$ readily falsifiable.

After publication of \tibet\ analysis \cite{Amenomori2021} several publications appeared  which suggested possible hadronic \cite{Dzhatdoev:2021xjh, Qiao:2021iua, Liu:2021lxk} or leptonic \cite{Fang:2021ylv} contributions to \tibet\ data. Main difference of our analysis from other approaches is that we consider constraints imposed by the Fermi/LAT data up to TeV energies and concentrate on the conservative pion-decay dominated model of emission from the Galactic disk. 

\section{Acknowledgments}
Work of A.N. and D.S. was supported in part by 
  the Ministry of science and higher education of Russian Federation under the contract 075-15-2020-778 in the framework of the Large scientific projects program within the national project ”Science”.
\bibliography{references}

\end{document}